\documentclass[11pt,english]{article}

\usepackage{lmodern}
\usepackage{color}

\usepackage[T1]{fontenc}
\usepackage[utf8x]{inputenc}
\usepackage{amsthm}
\usepackage{amsmath}
\usepackage{amssymb}
\usepackage{textcomp}
\usepackage{graphicx}
\usepackage{float}
\usepackage{a4wide}
\usepackage{enumerate}
\usepackage[parfill]{parskip}
\usepackage{todonotes}
\usepackage{enumitem}
\usepackage{relsize}

\newtheorem{rem}{Remark}

\title{Recipes for hedging exotics with illiquid vanillas\thanks{The authors would like to thank the Research Initiative ``Mod\'elisation des march\'es actions, obligations et d\'eriv\'es'' financed by HSBC France under the aegis of the Europlace Institute of Finance for their support regarding an early version of the paper. The authors would also like to thank Bastien Baldacci (Ecole Polytechnique), Philippe Bergault (Universit\'e Paris 1 Panth\'eon-Sorbonne), Fran\c{c}ois Bouscarle~(HSBC), Arnaud Gocsei~(HSBC), Nicolas Grandchamp des Raux~(HSBC), Greg Molin~~(HSBC), Jean Nguyen~(HSBC), and Jiang Pu (Institut Europlace de Finance) for the discussions they had on the topic. The readers should nevertheless be aware that the views, thoughts, and opinions expressed in the text belong solely to the authors.}}
\author{Joaquin Fernandez-Tapia\footnote{Institut Europlace de Finance. 28, place de la Bourse, 75002 Paris, France.}, Olivier Gu\'eant\footnote{Universit\'e Paris 1 Panth\'eon-Sorbonne. Centre d'Economie de la Sorbonne. 106, boulevard de l'H\^opital, 75013 Paris, France. Corresponding author. email: olivier.gueant@univ-paris1.fr}}
\date{}

\begin{document}

\maketitle

\begin{abstract}
In this paper, we address the question of the optimal Delta and Vega hedging of a book of exotic options when there are execution costs associated with the trading of vanilla options. In a framework where exotic options are priced using a market model (e.g.~a local volatility model recalibrated continuously to vanilla option prices) and vanilla options prices are driven by a stochastic volatility model, we show that, using simple approximations, the optimal dynamic Delta and Vega hedging strategies can be computed easily using variational techniques.
\end{abstract}

\section{Introduction}

The classical theory to price and manage derivatives contracts is based on the simplifying assumption of a frictionless market in which traded assets are liquid and agents incur no trading costs. Although this assumption is not realistic, it has led to very powerful pricing models that have been used on trading floors for more than 40 years. In spite of their success, classical pricing models need to be amended when reality is too far from classical modelling assumptions, in particular when there are transaction costs (e.g. a large bid-ask spread) or when the liquidity of the underlying assets is limited.\\

Classical option pricing models have been improved to account for transaction costs. One of the first models to include transaction costs is that of Leland \cite{tcleland} who proposed an amendment to the seminal model of Black and Scholes through a change in the volatility parameter to take account of both the transaction costs and the frequency of hedging. Several incomplete market approaches have been discussed in the literature for taking account of transaction costs. The super-replication approach, for instance, has been shown to be of no help to deal with transaction costs (see \cite{tccvitanic2, levantal, ssc}). Most authors introduced therefore utility functions to tackle the questions of the pricing and hedging of contingent claims in presence of transaction costs. Interesting examples include the paper of Barles and Soner \cite{tcbarlessoner} who obtained an elegant formula for the price using a modification of the implied volatility, the paper of Constantinides and Zariphopoulou \cite{cz} who obtained bounds on option prices or the paper of Cvitanic and Karatzas \cite{tccvitanic2}. Alternative approaches include also quantile hedging or the minimization of classical risk measures (see \cite{follmer1} and \cite{follmer2} for an introduction to these incomplete market methods).\\

In addition to the literature dealing with transaction costs, there is an interesting literature dealing with execution costs (and market impact). The classical paper by Çetin, Jarrow, and Protter \cite{scetin} (see also \cite{sbank} and \cite{scetin2,cetin3}) belongs to this category, although the authors do not phrase their approach in these terms. Their trader is not price-taker and the price she pays depends on the quantity she trades. Although it is very interesting, the main drawback of this framework is that it leads to prices identical to those of the Black-Scholes model. Çetin, Soner, and Touzi improved this approach in \cite{cetingamma} by adding a restriction to the space of admissible strategies (see also \cite{longstaff}), and they obtained positive liquidity costs and prices that depart from those of Black and Scholes. By considering absolutely continuous hedging strategies, a recent literature, inspired by the literature on optimal execution (see \cite{almgren, almgreninit} and~\cite{gueantgeneral}), obtained new results for the Delta hedging of options when liquidity has to be taken into account. Articles in this category include that of Rogers and Singh \cite{rogerssingh}, and the papers of Almgren and Li \cite{almgrenoption} and Guéant and Pu \cite{gueantoption} motivated by the observations of saw-tooth patterns on stocks the day of expiry of some options (see \cite{lehalle}).\footnote{See also  \cite{fplaten}, \cite{fschon}, and \cite{fsircar} for the feedback effect of hedging on option prices.}\\

All the previously discussed papers deal with Delta hedging when the underlying asset(s) is/are illiquid. When it comes to exotic equity derivatives, the hedging process does not only involve stocks or futures but rather stocks or futures and options, typically vanilla options. In this paper, we address the question of the optimal Delta and Vega hedging of a book of exotic derivatives under the assumption that the underlying asset is liquid but that trading vanilla options is costly. More precisely, we consider that exotic derivatives are valued using a market model and that vanilla options, whose price dynamics is driven by a stochastic volatility model, can be traded, with execution costs, using absolutely continuous strategies as in the literature on optimal execution.\\

In a mean-variance setting, using simple approximations, we manage to write the optimal hedging strategy of the trader as the solution of a deterministic variational problem involving three kinds of terms: terms to penalize fast execution (execution costs), terms modeling the Vega risk associated with the portfolio, and terms to profit from the trader's view on the market. When execution costs are quadratic as in the original paper of Almgren and Chriss (see~\cite{almgreninit}), our main result is a closed-form representation of the optimal hedging strategy in two different problems: one in which the portfolio is progressively hedged in the stochastic volatility model and another one in which we impose a complete unwinding of the risk in the market model while hedging in the stochastic volatility model.\\

The remainder of the text is organized as follows. In Section 2, we present the model, derive the equations for the profit and loss (PnL), and discuss the objective function of the trader. In Section 3, we present our simplifying approximations that make the problems tractable and we derive the closed-form solutions for the optimal hedging strategies of the trader in our two problems under these approximations.

\section{The model}

We consider a probability space $(\Omega,\mathcal{F},\mathbb{P}\big)$ with a filtration $(\mathcal{F}_{t})_{t\in \mathbb{R}^{+}}$ satisfying the usual conditions. Throughout the paper, we assume that all stochastic processes are defined on $\big(\Omega,\mathcal{F},(\mathcal{F}_t)_{t\in \mathbb{R}^{+}},\mathbb{P}\big)$.

\subsection{Market and price dynamics}

We consider an asset whose price dynamics is described by a one-factor stochastic volatility model of the form
\begin{align*}
\left\{
    \begin{array}{ll}
        dS_{t} = \mu_t S_t dt +\sqrt{\nu_{t}}S_{t}dW_{t}^{S,\mathbb{P}}  \\
        d\nu_{t}=a^{\mathbb{P}}(t,\nu_t)dt+\xi\sqrt{\nu_{t}}dW_{t}^{\nu,\mathbb{P}},
    \end{array}
\right.
\end{align*}
where $\xi \in \mathbb{R}^{+*}$, $(W_{t}^{S,\mathbb{P}},W_{t}^{\nu,\mathbb{P}})_{t\in \mathbb{R}^{+}}$ is a couple of Brownian motions with quadratic covariation given by $\rho=\frac{d\langle W^{S,\mathbb{P}},W^{\nu,\mathbb{P}}\rangle_t}{dt}  \in (-1,1)$, and
where the (bounded) adapted process $(\mu_t)_{t\in \mathbb{R}^{+}}$ and the function $a^{\mathbb{P}}$ are such that the processes are well defined (in particular, we assume that the process $(\nu_t)_{t\in \mathbb{R}^{+}}$ stays positive almost surely).\\

\begin{rem}
A classical example for the function $a^{\mathbb{P}}$ is that of the Heston model. In that case, $a^{\mathbb{P}}: (t,\nu) \mapsto \kappa^{\mathbb{P}}(\theta^{\mathbb{P}} - \nu)$ where $\kappa^{\mathbb{P}},\theta^{\mathbb{P}} \in \mathbb R^+$ satisfy the Feller condition $ 2 \kappa^{\mathbb{P}}\theta^{\mathbb{P}} > \xi^2$ (see~\cite{heston1993closed}).\\
\end{rem}

Assuming interest rates are equal to $0$, we introduce a risk-neutral / pricing probability measure $\mathbb{Q}$ equivalent to $\mathbb{P}$ under which the price and volatility processes become
\begin{align*}
\left\{
    \begin{array}{ll}
        dS_{t} = \sqrt{\nu_{t}}S_{t}dW_{t}^{S, \mathbb{Q}}  \\
        d\nu_{t}=a^{\mathbb{Q}}(t,\nu_t)dt+\xi\sqrt{\nu_{t}}dW_{t}^{\nu, \mathbb{Q}},
    \end{array}
\right.
\end{align*}
where $(W_{t}^{S, \mathbb{Q}},W_{t}^{\nu, \mathbb{Q}})_{t\in \mathbb{R}^{+}}$ is another couple of Brownian motions, this time under $\mathbb{Q}$, with quadratic covariation given by $\rho=\frac{d\langle W^{S, \mathbb{Q}},W^{\nu, \mathbb{Q}}\rangle_t}{dt}  \in (-1,1)$, and where $a^{\mathbb{Q}}$ is such that the processes are well defined.\\

We consider a set of $N\geq 1$ vanilla options written on the above underlying asset, and a book of exotic options whose value $P_t$ at time $t$ is assumed to be a function $\Pi$ of the time, the price of the underlying asset, and the price of the $N$ vanilla options.\\

Denoting the price of the $i$-th vanila option at time $t$ by $O_{t}^{i}$, we know that $O_t^{i} = \Omega^{i}(t,S_t,\nu_t)$ where $\Omega^{i}$ is solution of the following partial differential equation (PDE):
\begin{eqnarray*}
0 &=&  \frac{\partial \Omega^{i}}{\partial t} (t,S,\nu)   +  \frac{\partial \Omega^{i}}{\partial \nu} (t,S,\nu) a^{\mathbb{Q}}(t,\nu) \\
&+&  \frac{1}{2} \frac{\partial^2 \Omega^{i}}{\partial S^2} (t,S,\nu) \nu S^2 +  \frac{1}{2} \frac{\partial^2 \Omega^{i}}{\partial \nu^2} (t,S,\nu) \xi^2 \nu + \frac{\partial^2 \Omega^{i}}{\partial \nu \partial S} (t,S,\nu)\rho \xi \nu S,
\end{eqnarray*}
for $(t,S,\nu) \in [0,T^{i})\times {\mathbb{R}^{+}}^{2}$ where $T^i$ is the maturity of the $i$-th option.\\

Therefore, for $i \in \{1, \ldots, N\}$ and $t < T^i$, the dynamics of the $i$-th vanilla option is given by
\begin{eqnarray*}
&&dO_t^{i}\nonumber\\
 &=& \frac{\partial \Omega^{i}}{\partial t} (t,S_t,\nu_t) dt +  \frac{\partial \Omega^{i}}{\partial S} (t,S_t,\nu_t) dS_t + \frac{\partial \Omega^{i}}{\partial \nu} (t,S_t,\nu_t) d\nu_t \nonumber\\
&+& \frac{1}{2} \frac{\partial^2 \Omega^{i}}{\partial S^2} (t,S_t,\nu_t) d\langle S,S \rangle_t + \frac{1}{2} \frac{\partial^2 \Omega^{i}}{\partial \nu^2} (t,S_t,\nu_t) d\langle \nu,\nu \rangle_t + \frac{\partial^2 \Omega^{i}}{\partial \nu \partial S} (t,S_t,\nu_t) d\langle S,\nu \rangle_t \nonumber\\
&=&  \frac{\partial \Omega^{i}}{\partial t} (t,S_t,\nu_t) dt  + \frac{\partial \Omega^{i}}{\partial S} (t,S_t,\nu_t) \mu_t S_t dt +   \frac{\partial \Omega^{i}}{\partial S} (t,S_t,\nu_t)  \sqrt{\nu_t} S_t dW_t^{S,\mathbb{P}} \nonumber\\
&+&  \frac{\partial \Omega^{i}}{\partial \nu} (t,S_t,\nu_t) a^{\mathbb{P}}(t,\nu_t) dt +  \frac{\partial \Omega^{i}}{\partial \nu} (t,S_t,\nu_t)  \xi \sqrt{\nu_t} dW_t^{\nu,\mathbb{P}} \nonumber\\
&+&  \frac{1}{2} \frac{\partial^2 \Omega^{i}}{\partial S^2} (t,S_t,\nu_t) \nu_t S_t^2 dt +  \frac{1}{2} \frac{\partial^2 \Omega^{i}}{\partial \nu^2} (t,S_t,\nu_t) \xi^2 \nu_t dt + \frac{\partial^2 \Omega^{i}}{\partial \nu \partial S} (t,S_t,\nu_t)\rho \xi \nu_t S_t dt\nonumber\\
 &=& \left(  \frac{\partial \Omega^{i}}{\partial S} (t,S_t,\nu_t) \mu_t S_t +  \frac{\partial \Omega^{i}}{\partial \nu} (t,S_t,\nu_t) \left(a^{\mathbb{P}}(t,\nu_t) -a^{\mathbb{Q}}(t,\nu_t)\right) \right) dt \\
&+&  \frac{\partial \Omega^{i}}{\partial S} (t,S_t,\nu_t)  \sqrt{\nu_t} S_t dW_t^{S,\mathbb{P}} +  \frac{\partial \Omega^{i}}{\partial \nu} (t,S_t,\nu_t)  \xi \sqrt{\nu_t} dW_t^{\nu,\mathbb{P}}.\\
\end{eqnarray*}

The resulting dynamics for the value of the book of exotic options is
\begin{eqnarray*}
&&dP_t \nonumber\\
&=& \left(\frac{\partial \Pi}{\partial S}\left(t,S_t,O_t\right)  +  \sum_{i} \frac{\partial \Pi}{\partial O^{i}}\left(t,S_t,O_t\right) \frac{\partial \Omega^{i}}{\partial S} (t,S_t,\nu_t)  \right)  \sqrt{\nu}_t S_t  dW_t^{S,\mathbb{P}} \nonumber\\
&+& \sum_{i}  \frac{\partial \Pi}{\partial O^{i}}\left(t,S_t,O_t\right)\frac{\partial \Omega^{i}}{\partial \nu} (t,S_t,\nu_t)  \xi \sqrt{\nu_t} dW_t^{\nu,\mathbb{P}}\nonumber\\
&+& \frac{\partial \Pi}{\partial t}\left(t,S_t,O_t\right)  dt + \frac{\partial \Pi}{\partial S}\left(t,S_t,O_t\right)  \mu_t S_t dt \nonumber\\
&+&  \sum_{i} \frac{\partial \Pi}{\partial O^{i}}\left(t,S_t,O_t\right) \left(  \frac{\partial \Omega^{i}}{\partial S} (t,S_t,\nu_t) \mu_t S_t +  \frac{\partial \Omega^{i}}{\partial \nu} (t,S_t,\nu_t) \left(a^{\mathbb{P}}(t,\nu_t) -a^{\mathbb{Q}}(t,\nu_t)\right) \right)  dt \nonumber\\
&+& \frac{1}{2} \frac{\partial^2 \Pi}{\partial S^2}\left(t,S_t,O_t\right)  \nu_t S_t^2 dt + \frac{1}{2} \sum_{i}\sum_{i'} \frac{\partial^2 \Pi}{\partial O^{i}\partial O^{i'}}\left(t,S_t,O_t\right) \frac{\partial \Omega^{i}}{\partial \nu}(t,S_t,\nu_t)  \frac{\partial \Omega^{i'}}{\partial \nu} (t,S_t,\nu_t)  \xi^2 \nu_t dt\nonumber\\
&+& \sum_{i} \frac{\partial^2 \Pi}{\partial O^{i} \partial S} \left(t,S_t,O_t\right) \left( \frac{\partial \Omega^{i}}{\partial S} (t,S_t,\nu_t) \nu_t S_t^2  + \rho\frac{\partial \Omega^{i}}{\partial \nu} (t,S_t,\nu_t)  \xi \nu_t S_t \right) dt.\nonumber\\
\end{eqnarray*}

\subsection{Trading strategies and objective function}

We now consider a trader in charge of hedging the book of exotic options over a short period of time $[0,T]$ where $T$ is significantly smaller than $\min_i T^i$. For that purpose, she can trade the underlying asset with no friction but has to pay execution costs to trade the vanilla options. In what follows, we denote by $q_t^{S}$ the number of underlying assets held at time $t$ and by $q_t^{i}$ the number of $i$-th vanilla options held in the portfolio. For $i \in \{1, \ldots, N\}$, we assume that $dq_t^i = v^i_t dt$  and that the running costs paid (in addition to the MtM value) to trade the $i$-th option at velocity $v_t^i$ is given by $L^i(v_t^i)$ where $L^i$ satisfies the classical assumptions of execution costs functions:
\begin{itemize}
\item $L^i(0)=0$,
\item $L^i$ is increasing on $\mathbb{R}_{+}$ and decreasing on $\mathbb{R}_{-}$,
\item $L^i$ is strictly convex,
\item $L^i$ is asymptotically superlinear, that is $\lim_{\rho\to+\infty}\frac{L^i(\rho)}{\rho}  =  +\infty.$
\end{itemize}

The dynamics of the trader's PnL is therefore
\begin{eqnarray}
&&d \mathrm{PnL}_t\nonumber\\
&=& q_t^{S} dS_t + \sum_{i} q_t^{i} dO_t^{i} - \sum_{i} L^{i}(v_t^{i}) dt +  dP_t\nonumber\\
&=& q_t^{S} \mu_t S_t dt  + \sum_{i} q_t^{i} \left(  \frac{\partial \Omega^{i}}{\partial S} (t,S_t,\nu_t) \mu_t S_t +  \frac{\partial \Omega^{i}}{\partial \nu} (t,S_t,\nu_t) \left(a^{\mathbb{P}}(t,\nu_t) -a^{\mathbb{Q}}(t,\nu_t)\right) \right) dt \nonumber\\
&-& \sum_{i} L^{i}(v_t^{i}) dt \nonumber\\
&+& \left( \left(q_t^{S}+ \frac{\partial \Pi}{\partial S}\left(t,S_t,O_t\right)\right)  +  \sum_{i} \left(q_t^{i} + \frac{\partial \Pi}{\partial O^{i}}\left(t,S_t,O_t\right)\right) \frac{\partial \Omega^{i}}{\partial S} (t,S_t,\nu_t)  \right)  \sqrt{\nu}_t S_t  dW_t^{S,\mathbb{P}} \nonumber\\
&+& \sum_{i} \left(q_t^{i} + \frac{\partial \Pi}{\partial O^{i}}\left(t,S_t,O_t\right) \right)\frac{\partial \Omega^{i}}{\partial \nu} (t,S_t,\nu_t)  \xi \sqrt{\nu_t} dW_t^{\nu,\mathbb{P}}\nonumber\\
&+& \frac{\partial \Pi}{\partial t}\left(t,S_t,O_t\right)  dt + \frac{\partial \Pi}{\partial S}\left(t,S_t,O_t\right)  \mu_t S_t dt \nonumber\\
&+&  \sum_{i} \frac{\partial \Pi}{\partial O^{i}}\left(t,S_t,O_t\right) \left(  \frac{\partial \Omega^{i}}{\partial S} (t,S_t,\nu_t) \mu_t S_t +  \frac{\partial \Omega^{i}}{\partial \nu} (t,S_t,\nu_t) \left(a^{\mathbb{P}}(t,\nu_t) -a^{\mathbb{Q}}(t,\nu_t)\right) \right)  dt \nonumber\\
&+& \frac{1}{2} \frac{\partial^2 \Pi}{\partial S^2}\left(t,S_t,O_t\right)  \nu_t S_t^2 dt + \frac{1}{2} \sum_{i}\sum_{i'} \frac{\partial^2 \Pi}{\partial O^{i}\partial O^{i'}}\left(t,S_t,O_t\right) \frac{\partial \Omega^{i}}{\partial \nu}(t,S_t,\nu_t)  \frac{\partial \Omega^{i'}}{\partial \nu} (t,S_t,\nu_t)  \xi^2 \nu_t dt\nonumber\\
&+& \sum_{i} \frac{\partial^2 \Pi}{\partial O^{i} \partial S} \left(t,S_t,O_t\right) \left( \frac{\partial \Omega^{i}}{\partial S} (t,S_t,\nu_t) \nu_t S_t^2  + \rho\frac{\partial \Omega^{i}}{\partial \nu} (t,S_t,\nu_t)  \xi \nu_t S_t \right) dt. \label{pnl}
\end{eqnarray}

Ideally, we would like to maximize an objective function of the form $$\mathbb{E}\left[ \mathrm{PnL}_{T} \right] - \frac \gamma 2 \mathbb{V}\left[ \mathrm{PnL}_{T} \right]$$ over a constrained set of trading strategies.

Assuming that the trading strategies are such that the local martingales involved in \eqref{pnl} are martingales, we have that the part of $\mathbb{E}\left[ \mathrm{PnL}_{T} \right]$ that depends on $q^{S}$ and $(q^{i})_{i}$ is:
\begin{eqnarray*}
&&\mathbb{E}\left[ \mathlarger{\int}_0^{T} \left(q_t^{S} \mu_t S_t + \sum_{i} q_t^{i} \left(  \frac{\partial \Omega^{i}}{\partial S} (t,S_t,\nu_t) \mu_t S_t +  \frac{\partial \Omega^{i}}{\partial \nu} (t,S_t,\nu_t) \left(a^{\mathbb{P}}(t,\nu_t) -a^{\mathbb{Q}}(t,\nu_t)\right) \right) \right)dt \right]\\
&& - \mathbb{E}\left[ \mathlarger{\int}_0^{T} \sum_{i} L^{i}(v_t^{i}) dt\right].
\end{eqnarray*}

When it comes to the variance of the PnL, the computation is however far more cumbersome. For that reason, we consider an expansion of $ \mathbb{V}\left[ \mathrm{PnL}_{T} \right]$ for $T$ small:
\begin{eqnarray*}
&&\mathbb{V}\left[ \mathrm{PnL}_{T} \right]\nonumber\\
&=&\mathbb{V}\left[\mathlarger{\int}_0^{T} \left( \left(q_t^{S}+ \frac{\partial \Pi}{\partial S}\left(t,S_t,O_t\right)\right)  +  \sum_{i} \left(q_t^{i} + \frac{\partial \Pi}{\partial O^{i}}\left(t,S_t,O_t\right)\right) \frac{\partial \Omega^{i}}{\partial S} (t,S_t,\nu_t)  \right)  \sqrt{\nu}_t S_t  dW_t^{S,\mathbb{P}} \right.\nonumber\\
&+&\mathlarger{\int}_0^{T}\left. \sum_{i} \left(q_t^{i} + \frac{\partial \Pi}{\partial O^{i}}\left(t,S_t,O_t\right) \right)\frac{\partial \Omega^{i}}{\partial \nu} (t,S_t,\nu_t)  \xi \sqrt{\nu_t} dW_t^{\nu,\mathbb{P}}\right] + o(T)\nonumber\\
&=&\mathbb{E}\left[\mathlarger{\int}_0^{T} \left( \left(q_t^{S}+ \frac{\partial \Pi}{\partial S}\left(t,S_t,O_t\right)\right)  +  \sum_{i} \left(q_t^{i} + \frac{\partial \Pi}{\partial O^{i}}\left(t,S_t,O_t\right)\right) \frac{\partial \Omega^{i}}{\partial S} (t,S_t,\nu_t)  \right)^2  \nu_t S_t^2  dt\right]\nonumber\\
&+&\mathbb{E}\left[\mathlarger{\int}_0^{T} \left( \sum_{i} \left(q_t^{i} + \frac{\partial \Pi}{\partial O^{i}}\left(t,S_t,O_t\right) \right)\frac{\partial \Omega^{i}}{\partial \nu} (t,S_t,\nu_t) \right)^2 \xi^2 \nu_t dt\right]\nonumber\\
&+&2\rho \mathbb{E}\left[\mathlarger{\int}_0^{T} \left( \left(q_t^{S}+ \frac{\partial \Pi}{\partial S}\left(t,S_t,O_t\right)\right)  +  \sum_{i} \left(q_t^{i} + \frac{\partial \Pi}{\partial O^{i}}\left(t,S_t,O_t\right)\right) \frac{\partial \Omega^{i}}{\partial S} (t,S_t,\nu_t)  \right)\right.\nonumber\\
&&\left.\times\left( \sum_{i} \left(q_t^{i} + \frac{\partial \Pi}{\partial O^{i}}\left(t,S_t,O_t\right) \right)\frac{\partial \Omega^{i}}{\partial \nu} (t,S_t,\nu_t) \right)\xi\nu_t S_t dt\right] + o(T).\nonumber
\end{eqnarray*}

Therefore, to the first order in $T$, we can approximate our problem by the minimization, over $q^{S}$ and $(v^{i})_{i}$  in a set to be specified,  of
\begin{eqnarray*}
&&\mathbb{E}\left[ - \mathlarger{\int}_0^{T} \left(q_t^{S} \mu_t S_t + \sum_{i} q_t^{i} \left(  \frac{\partial \Omega^{i}}{\partial S} (t,S_t,\nu_t) \mu_t S_t +  \frac{\partial \Omega^{i}}{\partial \nu} (t,S_t,\nu_t) \left(a^{\mathbb{P}}(t,\nu_t) -a^{\mathbb{Q}}(t,\nu_t)\right) \right) \right)dt \right]\nonumber\\
&+& \mathbb{E}\left[ \mathlarger{\int}_0^{T} \sum_{i} L^{i}(v_t^{i}) dt\right]\nonumber\\
&+& \frac 12  \gamma \left[ \mathbb{E}\left[\mathlarger{\int}_0^{T} \left( \left(q_t^{S}+ \frac{\partial \Pi}{\partial S}\left(t,S_t,O_t\right)\right)  +  \sum_{i} \left(q_t^{i} + \frac{\partial \Pi}{\partial O^{i}}\left(t,S_t,O_t\right)\right) \frac{\partial \Omega^{i}}{\partial S} (t,S_t,\nu_t)  \right)^2  \nu_t S_t^2  dt\right]\right.\nonumber\\
&+&\mathbb{E}\left[\mathlarger{\int}_0^{T} \left( \sum_{i} \left(q_t^{i} + \frac{\partial \Pi}{\partial O^{i}}\left(t,S_t,O_t\right) \right)\frac{\partial \Omega^{i}}{\partial \nu} (t,S_t,\nu_t) \right)^2 \xi^2 \nu_t dt\right]\nonumber\\
&+&2\rho\mathbb{E}\left[\mathlarger{\int}_0^{T} \left( \left(q_t^{S}+ \frac{\partial \Pi}{\partial S}\left(t,S_t,O_t\right)\right)  +  \sum_{i} \left(q_t^{i} + \frac{\partial \Pi}{\partial O^{i}}\left(t,S_t,O_t\right)\right) \frac{\partial \Omega^{i}}{\partial S} (t,S_t,\nu_t)  \right)\right.\nonumber\\
&&\left.\left.\times\left( \sum_{i} \left(q_t^{i} + \frac{\partial \Pi}{\partial O^{i}}\left(t,S_t,O_t\right) \right)\frac{\partial \Omega^{i}}{\partial \nu} (t,S_t,\nu_t) \right)\xi\nu_t S_t dt\right]\right].\\
\end{eqnarray*}

Denoting $$u_t = \left( \left(q_t^{S}+ \frac{\partial \Pi}{\partial S} (t,S_t,O_t) \right)  +  \sum_{i} \left(q_t^{i} + \frac{\partial \Pi}{\partial O^i}(t,S_t,O_t)  \right) \frac{\partial \Omega^{i}}{\partial S} (t,S_t,\nu_t)  \right) \sqrt{\nu_t}  S_t,$$ our problem boils down to minimizing over  $u$ and $(v^{i})_{i}$  in a set to be specified the expression
\begin{eqnarray*}
&&\mathbb{E}\left[ \mathlarger{\int}_0^{T} \left(- \mu_t \frac{u_t}{\sqrt{\nu_t}} + \mu_t S_t \left( \frac{\partial \Pi}{\partial S}\left(t,S_t,O_t\right) + \sum_{i} \frac{\partial \Pi}{\partial O^i}(t,S_t,O_t) \frac{\partial \Omega^{i}}{\partial S} (t,S_t,\nu_t) \right) \right) dt\right]\nonumber\\
&+& \mathbb{E}\left[ \mathlarger{\int}_0^{T}  \left(- \sum_{i} q_t^{i} \left( \frac{\partial \Omega^{i}}{\partial \nu} (t,S_t,\nu_t) \left(a^{\mathbb{P}}(t,\nu_t) -a^{\mathbb{Q}}(t,\nu_t)\right) \right) + \sum_{i} L^{i}(v_t^{i}) \right)dt \right]\nonumber\\
&+& \frac 12  \gamma \mathbb{E}\left[\mathlarger{\int}_0^{T}\left( u_t^2 +  \left( \sum_{i} \left(q_t^{i} + \frac{\partial \Pi}{\partial O^i}\left(t,S_t,O_t\right) \right) \frac{\partial \Omega^{i}}{\partial \nu} (t,S_t,\nu_t)  \right)^2 \xi^2 \nu_t\right.\right.\nonumber\\
&&\qquad\left.\left. \vphantom{u_t^2 +  \left( \sum_{i} \left(q_t^{i} + \frac{\partial \Pi}{\partial O^i}\left(t,S_t,O_t\right)\right) \frac{\partial \Omega^{i}}{\partial \nu} (t,S_t,\nu_t)  \right)^2 \xi^2 \nu_t} \quad  +\ 2\rho u_t  \sum_{i} \left(q_t^{i} + \frac{\partial \Pi}{\partial O^i}\left(t,S_t,O_t\right) \right)  \frac{\partial \Omega^{i}}{\partial \nu} (t,S_t,\nu_t) \xi \sqrt{\nu_t} \right)dt \right].\\
\end{eqnarray*}

If there is no constraint on the process $(u_t)_t$, and if we consider that the process $(q_t)_t = (q_t^{1}, \ldots, q_t^{N})_t$ is given, then the optimal value of $u_t$ must verify
$$u_t^* = \frac{\mu_t}{\gamma \sqrt{\nu_t}}- \rho  \xi \sqrt{\nu_t} \sum_{i} \left(q_t^{i} +  \frac{\partial \Pi}{\partial O^i}\left(t,S_t,O_t\right)  \right)  \frac{\partial \Omega^{i}}{\partial \nu} (t,S_t,\nu_t),$$
i.e.
\begin{equation*}
q_t^{S*} =  \frac{\mu_t}{\gamma \nu_t S_t} - \left( \frac{\partial \Pi}{\partial S} (t,S_t,O_t) +  \sum_{i} \left(q_t^{i} +  \frac{\partial \Pi}{\partial O^i}\left(t,S_t,O_t\right) \right) \left(\frac{\partial \Omega^{i}}{\partial S} (t,S_t,\nu_t) + \frac{\rho\xi}{S_t} \frac{\partial \Omega^{i}}{\partial \nu} (t,S_t,\nu_t)  \right)\right).
\end{equation*}
This formula has two components: (i) the term $\frac{\mu_t}{\gamma \nu_t S_t}$ that corresponds to the optimal number of underlying assets to hold in a pure dynamic mean-variance portfolio choice problem \`a la Merton, given the trader's view (see \cite{merton} for instance), and (ii) a Delta term that corresponds to the Delta of the portfolio in our model with a correction term taking account of the possibility to Delta-hedge part of the Vega in the stochastic volatility model whenever the vol-spot correlation parameter $\rho$ is not equal to $0$.\\

Now, plugging $u_t^*$ in the optimization problem, we see that our objective function writes
\begin{eqnarray*}
&&\mathbb{E}\left[ \mathlarger{\int}_0^{T} \mu_t S_t \left( \frac{\partial \Pi}{\partial S}\left(t,S_t,O_t\right) + \sum_{i} \frac{\partial \Pi}{\partial O^i}(t,S_t,O_t) \frac{\partial \Omega^{i}}{\partial S} (t,S_t,\nu_t) \right) dt\right]\nonumber\\
&+& \mathbb{E}\left[ \mathlarger{\int}_0^{T}  \left(- \sum_{i} q_t^{i} \left( \frac{\partial \Omega^{i}}{\partial \nu} (t,S_t,\nu_t) \left(a^{\mathbb{P}}(t,\nu_t) -a^{\mathbb{Q}}(t,\nu_t)\right) \right) + \sum_{i} L^{i}(v_t^{i}) \right)dt \right]\nonumber\\
&+&\mathbb{E}\left[\mathlarger{\int}_0^{T}\left(- \frac{\mu_t^2}{2\gamma \nu_t} + \frac 12  \gamma (1-\rho^2)\left( \sum_{i} \left(q_t^{i} + \frac{\partial \Pi}{\partial O^i}\left(t,S_t,O_t\right) \frac{\partial \Omega^{i}}{\partial \nu} (t,S_t,\nu_t)  \right)  \right)^2 \xi^2 \nu_t\right.\right.\nonumber\\
&&\qquad\left.\left. \vphantom{- \frac{\mu_t^2}{2\gamma \nu_t} + \frac 12  \gamma (1-\rho^2)\left( \sum_{i} \left(q_t^{i} + \frac{\partial \Pi}{\partial O^i}\left(t,S_t,O_t\right) \right) \frac{\partial \Omega^{i}}{\partial \nu} (t,S_t,\nu_t)  \right)^2 \xi^2 \nu_t} +\ \rho \mu_t \xi  \sum_{i} \left(q_t^{i} + \frac{\partial \Pi}{\partial O^i}\left(t,S_t,O_t\right) \right)  \frac{\partial \Omega^{i}}{\partial \nu} (t,S_t,\nu_t) \right)dt \right].\nonumber
\end{eqnarray*}
or, up to additive terms independent of the trading strategies,
\begin{align}
&\mathbb{E}\left[ \mathlarger{\int}_0^{T}  \left(- \sum_{i} \left(q_t^{i} + \frac{\partial \Pi}{\partial O^i}\left(t,S_t,O_t\right) \right) \left( \frac{\partial \Omega^{i}}{\partial \nu} (t,S_t,\nu_t) \left(a^{\mathbb{P}}(t,\nu_t) -a^{\mathbb{Q}}(t,\nu_t)\right) \right) + \sum_{i} L^{i}(v_t^{i}) \right)dt \right]\nonumber\\
&+\mathbb{E}\left[\mathlarger{\int}_0^{T}\left(\frac 12  \gamma (1-\rho^2)\left( \sum_{i} \left(q_t^{i} + \frac{\partial \Pi}{\partial O^i}\left(t,S_t,O_t\right) \right) \frac{\partial \Omega^{i}}{\partial \nu} (t,S_t,\nu_t)  \right)^2 \xi^2 \nu_t\right.\right.\nonumber\\
&\qquad\left.\left. \vphantom{- \frac{\mu_t^2}{2\gamma \nu_t} + \frac 12  \gamma (1-\rho^2)\left( \sum_{i} \left(q_t^{i} + \frac{\partial \Pi}{\partial O^i}\left(t,S_t,O_t\right)\right) \frac{\partial \Omega^{i}}{\partial \nu} (t,S_t,\nu_t)    \right)^2 \xi^2 \nu_t} +\ \rho \mu_t \xi  \sum_{i} \left(q_t^{i} + \frac{\partial \Pi}{\partial O^i}\left(t,S_t,O_t\right) \right)  \frac{\partial \Omega^{i}}{\partial \nu} (t,S_t,\nu_t) \right)dt \right].\label{pbm_gen_0}
\end{align}
where $(q_t)_t = (q_t^{1}, \ldots, q_t^N)_t$ has to satisfy constraints to be specified.\\

\section{Towards variational problems}

\subsection{Simplifying approximations}

Although considering a short-time horizon simplifies the objective function, the problem remains difficult to address numerically because of its dimensionality.  We consider therefore an approximation of the above optimization problem where the terms characterizing the dynamics of the underlying asset and the Vega terms are freezed over $[0,T]$. More precisely, we consider the following approximations:\footnote{Similar approximations are used in the recently published paper \cite{balda} to address an option market making problem.}
\begin{itemize}
\item The Sharpe ratio $\frac{\mu_t}{\sqrt{\nu_t}}$ is approximated by a constant denoted by $\frak{s}$.
\item The rescaled difference between the drifts of the volatility under $\mathbb{P}$ and $\mathbb{Q}$, defined as $\frac{a^{\mathbb{P}}(t,\nu_t) -a^{\mathbb{Q}}(t,\nu_t)}{\sqrt{\nu_t}}$, is approximated by a constant denoted by  $\zeta$.
\item For each $i\in \{1, \ldots, N\}$, the Vega of the $i$-th option in the stochastic volatility model, i.e. $\frac{\partial \Omega^{i}}{\partial \sqrt{\nu}} (t,S_t,\nu_t) = 2 \sqrt{\nu_t} \frac{\partial \Omega^{i}}{\partial \nu} (t,S_t,\nu_t)$ ($i \in \{1, \ldots, N\}$), is approximated by a constant denoted by $\mathcal{V}_{\textrm{SV}}^{i}$.
\item The Vegas of the exotic portfolio in the market model (with respect the $N$ implied volatilities) are constant and equal to $\mathcal{V}_{\textrm{MM}}^{i}$ ($i \in \{1, \ldots, N\}$). Therefore, for each $i\in \{1, \ldots, N\}$, $\frac{\partial \Pi}{\partial O^{i}} (t,S_t,O_t)$ is equal to $\frac{\mathcal{V}_{\textrm{MM}}^{i}}{\mathcal{V}_{\textrm{BS}}^{i}}$: the ratio of the Vega of the book of exotic options with respect to the implied volatility associated with the $i$-th vanilla option in the market model and the Vega of that vanilla option in the Black-Scholes model.\\
\end{itemize}

With these approximations, \eqref{pbm_gen_0} writes
\begin{eqnarray*}
&& \mathbb{E}\left[ \mathlarger{\int}_0^{T}  \left(\sum_{i} L^{i}(v_t^{i}) + \frac 18  \gamma (1-\rho^2) \xi^2\left( \sum_{i} \left(q_t^{i} + \frac{\mathcal{V}_{\textrm{MM}}^{i}}{\mathcal{V}_{\textrm{BS}}^{i}} \right) \mathcal{V}_{\textrm{SV}}^{i} \right)^2 \right)dt \right]\nonumber\\
&+&\mathbb{E}\left[\mathlarger{\int}_0^{T} \frac 12 \left(\rho \frak{s} \xi - \zeta\right)  \sum_{i} \left(q_t^{i} + \frac{\mathcal{V}_{\textrm{MM}}^{i}}{\mathcal{V}_{\textrm{BS}}^{i}} \right)  \mathcal{V}_{\textrm{SV}}^{i} dt \right].\\
\end{eqnarray*}

Focusing on deterministic strategies (that can be shown to be optimal by following the same reasoning as in \cite{schied} -- see also \cite{gueantgeneral}) our objective function (for minimizing) is in fact:
\begin{eqnarray*}
&&\mathlarger{\int}_0^{T}  \left(\sum_{i} L^{i}(v_t^{i}) + \frac 18  \gamma (1-\rho^2) \xi^2\left( \sum_{i} \left(q_t^{i} + \frac{\mathcal{V}_{\textrm{MM}}^{i}}{\mathcal{V}_{\textrm{BS}}^{i}} \right) \mathcal{V}_{\textrm{SV}}^{i} \right)^2 \right)dt\nonumber\\
&+&\mathlarger{\int}_0^{T} \frac 12 \left(\rho \frak{s} \xi - \zeta\right)  \sum_{i} \left(q_t^{i} + \frac{\mathcal{V}_{\textrm{MM}}^{i}}{\mathcal{V}_{\textrm{BS}}^{i}} \right)  \mathcal{V}_{\textrm{SV}}^{i} dt.\\
\end{eqnarray*}

\subsection{Two hedging problems}

In what follows, we consider two different problems corresponding to two different sets of constraints for the trading strategies. In the first problem, we simply impose the initial state of the portfolio of vanilla options, i.e. $q_0$ is given. In the second one, we additionally impose that the portfolio of vanilla options at time $T$  makes the portfolio hedged in the market model (up to the approximation of constant Vegas), i.e.  $q_T = -\frak v$ where $\frak{v} :=  \left(\frac{\mathcal{V}_{\textrm{MM}}^{1}}{\mathcal{V}_{\textrm{BS}}^{1}}, \ldots,  \frac{\mathcal{V}_{\textrm{MM}}^{N}}{\mathcal{V}_{\textrm{BS}}^{N}}\right)'$. In other words, the first problem is a Vega hedging problem in the stochastic volatility model while the second is a Vega Bucket cancellation problem with Vega hedging in the stochastic volatility model that corresponds better to the real problem faced by traders.\\

\begin{rem}
Because of the above approximations, our optimization problems are only meaningful over a short period of time. This may be regarded as a problem but it must be noted that one can use the output of our models over a short period of time and then run the model again with updated values of $\frak s$, $\zeta$, and the Vegas. Although this approach is time-inconsistent, it is a classical practice in applied optimal control, when parameters are estimated online for instance.
\end{rem}

\paragraph{Vega hedging in the stochastic volatility model}

In the first problem we consider, we simply impose the initial condition (i.e. $q_0$ given). Therefore, the problem writes
\begin{align*}
\inf_{(q_t)_t, q_0 \text{ given}}\mathlarger{\int}_0^{T} & \left(\sum_{i} L^{i}(v_t^{i}) + \frac 18  \gamma (1-\rho^2) \xi^2\left( \sum_{i} \left(q_t^{i} + \frac{\mathcal{V}_{\textrm{MM}}^{i}}{\mathcal{V}_{\textrm{BS}}^{i}} \right) \mathcal{V}_{\textrm{SV}}^{i} \right)^2 \right)dt\nonumber\\
&+\mathlarger{\int}_0^{T} \frac 12 \left(\rho \frak{s} \xi - \zeta\right)  \sum_{i} \left(q_t^{i} + \frac{\mathcal{V}_{\textrm{MM}}^{i}}{\mathcal{V}_{\textrm{BS}}^{i}} \right)  \mathcal{V}_{\textrm{SV}}^{i} dt.
\end{align*}

This problem is a problem of Bolza and we know, for instance from the work of Rockafellar \cite{rockafellar}  (see also \cite{cannarsa}), that there is a unique optimal trajectory $ t \in[0,T] \mapsto q^*(t) = (q^{*1}(t), \ldots, q^{*N}(t))'$ characterized by the following Hamiltonian system:
\begin{equation}
\label{Hamiltonian}
\begin{cases}
\dot p(t) = \frac 14 \gamma (1-\rho^2) \xi^2 \mathcal{V}_{\textrm{SV}} \mathcal{V}_{\textrm{SV}}' (q^*(t) + \frak{v}) + \frac 12 \left(\rho \frak{s} \xi - \zeta\right) \mathcal{V}_{\textrm{SV}} , \qquad p(T) = 0, \\
\dot{q}^{*i}(t) = {H^{i}}'(p^i(t)), \qquad q^*(0) = q_0 , \\
\end{cases}
\end{equation}
where the Hamiltonian functions $(H^1, \ldots, H^N)$ are defined by
$$\forall i \in \{1, \ldots, N\}, \quad H^i(z) = \sup_{v} vz - L^i(v).$$

In the case where execution cost functions are quadratic as in the original paper of Almgren and Chriss \cite{almgreninit}, we can in fact solve in closed form the system \eqref{Hamiltonian}. If indeed we have $L^i(v) = \eta^i v^2 $ for all $i \in \{1, \ldots, N\}$, then  \eqref{Hamiltonian} is the linear system
\begin{equation*}
\begin{cases}
\dot p(t) = \frac 14 \gamma (1-\rho^2) \xi^2 \mathcal{V}_{\textrm{SV}} \mathcal{V}_{\textrm{SV}}' (q^*(t) + \frak{v}) + \frac 12 \left(\rho \frak{s} \xi - \zeta\right) \mathcal{V}_{\textrm{SV}}, \qquad p(T) = 0, \\
\dot{q}^*(t) =\frac 1{2} \Lambda p(t), \qquad q^*(0) =  q_0, \\
\end{cases}
\end{equation*}
or equivalently
\begin{equation}
\label{edo2}
\ddot{q}^*(t) = \frac 18 \gamma (1-\rho^2) \xi^2 \Lambda\mathcal{V}_{\textrm{SV}} \mathcal{V}_{\textrm{SV}}'q^*(t) + \frac 18 \gamma (1-\rho^2) \xi^2 \Lambda\mathcal{V}_{\textrm{SV}} \mathcal{V}_{\textrm{SV}}'\frak v + \frac 12 \left(\rho \frak{s} \xi - \zeta\right) \Lambda\mathcal{V}_{\textrm{SV}}
\end{equation}
with boundary conditions $q^*(0) =  q_0$ and  $\dot{q}^*(T) = 0$, where $\Lambda = \begin{pmatrix}\frac 1{\eta^{1}} & & \\ & \ddots & \\ & & \frac{1}{\eta^{N}}\end{pmatrix}$.\\

In the case of quadratic execution cost functions, the problem therefore boils to a linear ordinary differential equation of order 2 that can be addressed using standard techniques.\\

For solving this ordinary differential equation, it is interesting to notice that $$\forall t \in [0,T], \ddot{q}^*(t) \in \textrm{span}( \Lambda\mathcal{V}_{\textrm{SV}}).$$ As $\dot{q}^*(T) = 0$, we deduce that $$\forall t \in [0,T], \dot{q}^*(t) \in \textrm{span}( \Lambda\mathcal{V}_{\textrm{SV}}).$$
Since $q^*(0) =  q_0$, there exists a $C^2$ function $\alpha$ such that $\forall t \in [0,T], q^*(t) = q_0 + \alpha(t) \Lambda\mathcal{V}_{\textrm{SV}}$.\\

Using \eqref{edo2}, we obtain
\begin{equation*}
\ddot{\alpha}(t) = \frac 18 \gamma (1-\rho^2) \xi^2  \mathcal{V}_{\textrm{SV}}'\Lambda\mathcal{V}_{\textrm{SV}} \alpha(t) + \frac 18 \gamma (1-\rho^2) \xi^2 \mathcal{V}_{\textrm{SV}}'(\frak v + q_0) + \frac 12 \left(\rho \frak{s} \xi - \zeta\right),
\end{equation*}
with boundary conditions $\alpha(0) =  0$ and $\dot{\alpha}(T) = 0$.\\

It is then straightforward to see that
\begin{equation}
\label{alpha}
\forall t \in [0,T], \alpha(t) = -\left(\frac{\mathcal{V}_{\textrm{SV}}'(\frak v + q_0)}{\mathcal{V}_{\textrm{SV}}'\Lambda\mathcal{V}_{\textrm{SV}}} + \frac{\rho \frak{s} \xi - \zeta}{2\lambda}\right) \left( 1 - \frac{\cosh(\sqrt{\lambda}(T-t))}{\cosh(\sqrt{\lambda}T)} \right),
\end{equation}
where $\lambda = \frac 18 \gamma (1-\rho^2) \xi^2  \mathcal{V}_{\textrm{SV}}'\Lambda\mathcal{V}_{\textrm{SV}}$, and therefore
\begin{equation*}
\forall t \in [0,T], q^*(t) = q_0 - \left(\frac{\mathcal{V}_{\textrm{SV}}'(\frak v+ q_0)}{\mathcal{V}_{\textrm{SV}}'\Lambda\mathcal{V}_{\textrm{SV}}} + \frac{\rho \frak{s} \xi - \zeta}{2\lambda}\right) \left(1- \frac{\cosh(\sqrt{\lambda}(T-t))}{\cosh(\sqrt{\lambda}T)} \right) \Lambda \mathcal{V}_{\textrm{SV}}.
\end{equation*}

This result deserves to be commented upon.\\

First, in this optimization problem, the optimal Vega hedging strategy in the case of quadratic execution costs always consists in trading the same basket of vanilla options. This result may seem odd at first sight but one has to remember that it is based on a one-factor dynamics for the vanilla options and that, because there is no final constraint in this first problem, the trader only hedges her position in the stochastic volatility model. The weights of the basket are given (up to a multiplicative constant) by the vector $\Lambda \mathcal{V}_{\textrm{SV}} = \left(\frac{ \mathcal{V}_{\textrm{SV}}^1}{\eta^1}, \ldots, \frac{ \mathcal{V}_{\textrm{SV}}^N}{\eta^N}\right)'$: the higher its Vega and the more liquid a vanilla option, the higher its weight in the basket.  It is in particular important to notice that the basket of vanilla options is independent of the portfolio of exotic options.\\

Second, the volume we trade of that basket of vanilla options depends on the Vegas of the portfolio of exotic options (see \eqref{alpha}). For instance, in the case where the trader has no view on the market (i.e. $\frak{s} = 0$ and $\zeta = 0$), whether the trader should buy or sell the basket of vanilla options depends on the scalar product $\mathcal{V}_{\textrm{SV}}'\frak v$ of the Vegas of the vanilla options in the stochastic volatility model and the sensitivities of the portfolio of exotic options to the different vanilla options: if $\mathcal{V}_{\textrm{SV}}'\frak v$ is positive (resp. negative) the trader will sell (resp. buy) the basket of vanilla options in order to hedge the portfolio of exotic options. Interestingly, the composition of the portfolio of exotic options determines the level (i.e. the scale) of the trading curve but not the shape (as a function of time) -- which only depends on the constant $\lambda$.\\

Third, the trader's view on the market influences her strategy. Indeed, even if the portfolio is already hedged in the market model, i.e. $q_0 = -\frak v$, the trader trades whenever $\rho\xi\frak s - \zeta$ is not equal to $0$. The trader's view on the market impacts the trading strategy through two effects. A first effect is related to the trader's view on the dynamics of the instantaneous volatility in the stochastic volatility model: the larger $\zeta$, i.e. the more upward the view of the trader on the instantaneous volatility, the bigger the incentive to buy the basket of vanilla options. Since the Vegas of the vanilla options in the stochastic volatility model are positive, this means that the more upward the view of the trader on the instantaneous volatility, the bigger the incentive to buy vanilla options. This was expected. The second effect is more subtle: the higher the product of the vol-spot correlation in the stochastic volatility model and the view on the Sharpe ratio of the underlying asset, the smaller the incentive to buy the basket of vanilla options (and the vanilla options themselves because of the sign the Vegas in the stochastic volatility model). This effect is due to the incentive of using the underlying asset to partially hedge the Vega of the portfolio in the stochastic volatility model when the vol-spot correlation is not equal to nought. To better understand this effect, let us consider the case where $\rho$ and $\frak{s}$ are positive and let us consider the case $\rho = \frak s = 0$ as a benchmark case. Would the trader buy more of the basket of vanilla options in the former case than in the latter (benchmark) case, then, because the vol-spot correlation is positive, the trader would sell more of the underlying asset in order to Delta-hedge part of the Vega of the portfolio in the stochastic volatility model, and this would result in an expected loss as the Sharpe ratio is positive. Subsequently, there is a reduced incentive to buy the basket of vanilla options when $\rho$ and $\frak{s}$ are positive.\\

\paragraph{Vega Bucket cancellation with Vega hedging in the stochastic volatility model}

In practice, the above problem does not lead to a complete cancellation of the Vega risk exposure in the market model. To reach this objective, we consider our second problem
\begin{align*}
\inf_{(q_t)_t, q_0 \text{ given}, q_T = -\frak v}\mathlarger{\int}_0^{T} & \left(\sum_{i} L^{i}(v_t^{i}) + \frac 18  \gamma (1-\rho^2) \xi^2\left( \sum_{i} \left(q_t^{i} + \frac{\mathcal{V}_{\textrm{MM}}^{i}}{\mathcal{V}_{\textrm{BS}}^{i}} \right) \mathcal{V}_{\textrm{SV}}^{i} \right)^2 \right)dt\nonumber\\
&+\mathlarger{\int}_0^{T} \frac 12 \left(\rho \frak{s} \xi - \zeta\right)  \sum_{i} \left(q_t^{i} + \frac{\mathcal{V}_{\textrm{MM}}^{i}}{\mathcal{V}_{\textrm{BS}}^{i}} \right)  \mathcal{V}_{\textrm{SV}}^{i} dt
\end{align*}
in which the trader hedges her portfolio in the stochastic volatility model, which describes the dynamics of vanilla options, and has to reach at time $T$ a position that hedges her portfolio in the market model.

This problem is a problem of Bolza and, as above, there is a unique optimal trajectory $ t \in[0,T] \mapsto q^*(t) = (q^{*1}(t), \ldots, q^{*N}(t))'$ characterized by the following Hamiltonian system:
\begin{equation}
\label{Hamiltonian_2}
\begin{cases}
\dot p(t) = \frac 14 \gamma (1-\rho^2) \xi^2 \mathcal{V}_{\textrm{SV}} \mathcal{V}_{\textrm{SV}}' (q^*(t) + \frak{v}) + \frac 12 \left(\rho \frak{s} \xi - \zeta\right) \mathcal{V}_{\textrm{SV}}\\
\dot{q}^{*i}(t) = {H^{i}}'(p^i(t)), \qquad q^*(0) = q_0 , q^*(T) = -\frak v.\\
\end{cases}
\end{equation}

In the case where execution cost functions are quadratic as above, \eqref{Hamiltonian_2} is in fact the linear system
\begin{equation*}
\begin{cases}
\dot p(t) = \frac 14 \gamma (1-\rho^2) \xi^2 \mathcal{V}_{\textrm{SV}} \mathcal{V}_{\textrm{SV}}' (q^*(t) + \frak{v}) + \frac 12 \left(\rho \frak{s} \xi - \zeta\right) \mathcal{V}_{\textrm{SV}},\\
\dot{q}^*(t) =\frac 1{2} \Lambda p(t), \qquad q^*(0) = q_0, q^*(T) = -\frak v, \\
\end{cases}
\end{equation*}
or equivalently
\begin{equation}
\label{edo2_2}
\ddot{q}^*(t) = \frac 18 \gamma (1-\rho^2) \xi^2 \Lambda\mathcal{V}_{\textrm{SV}} \mathcal{V}_{\textrm{SV}}'q^*(t) + \frac 18 \gamma (1-\rho^2) \xi^2 \Lambda\mathcal{V}_{\textrm{SV}} \mathcal{V}_{\textrm{SV}}'\frak v + \frac 12 \left(\rho \frak{s} \xi - \zeta\right) \Lambda\mathcal{V}_{\textrm{SV}}
\end{equation}
with boundary conditions $q^*(0) =  q_0$ and  $q^*(T) = -\frak v$.\\

As above, in the case of quadratic execution cost functions, the problem therefore boils to a linear ordinary differential equation of order 2 that can be addressed using standard techniques.\\

As above, we notice that
$$ \forall t \in [0,T], \ddot{q}^{*}(t) \in \text{span}(\Lambda \mathcal{V}_{\textrm{BS}}).$$ Then, recalling that
\begin{eqnarray*}
\forall t \in [0,T], q^{*}(t) &=& \frac{(T-t) q(0) + t q(T)}{T} + \int_0^t \ddot{q}^{*}(s)(t-s) ds - \frac t T \int_0^T \ddot{q}^{*}(s)(T-s) ds\\
 &=& \left(1- \frac tT\right)q_0-\frac t T\frak v + \int_0^t \ddot{q}^{*}(s)(t-s) ds - \frac t T \int_0^T \ddot{q}^{*}(s)(T-s) ds,
 \end{eqnarray*}
we clearly see that there exists a $C^2$ function $\alpha$ such that
$$\forall t \in [0,T], q^{*}(t) = \left(1- \frac tT\right)q_0-\frac t T\frak v + \alpha(t) \Lambda \mathcal{V}_{\textrm{BS}}.$$

From \eqref{edo2_2}, we deduce that
$$\forall t \in [0,T],  \ddot{\alpha}(t) = \lambda \alpha(t) + \frac 18 \gamma (1-\rho^2) \xi^2 \left( 1 - \frac{t}T\right) \mathcal{V}_{\textrm{SV}}'(\frak v + q_0) + \frac 12 \left(\rho \frak{s} \xi - \zeta\right) ,$$ with boundary conditions $\alpha(0) = 0$ and $\alpha(T) = 0$.

It is then straightforward to see that
\begin{eqnarray*}\forall t \in [0,T], \alpha(t) &=& \frac{\mathcal{V}_{\textrm{SV}}'(\frak{v} + q_0)}{\mathcal{V}_{\textrm{SV}}'\Lambda \mathcal{V}_{\textrm{SV}}}\left( \frac{\sinh(\sqrt{\lambda}(T-t))}{\sinh(\sqrt{\lambda}T)} - \left(1-\frac tT\right)\right)\\
&+& \frac{\rho \frak{s} \xi - \zeta}{2\lambda} \left( \frac{\sinh(\sqrt{\lambda}t) + \sinh(\sqrt{\lambda}(T-t))}{\sinh(\sqrt{\lambda}T)} - 1\right).\\
\end{eqnarray*}

The optimal hedging strategy in this second problem, which is more in line with the actual problem faced by the trader, is then
\begin{eqnarray*}\forall t \in [0,T], q^{*}(t) &=& \left(1- \frac tT\right)q_0-\frac t T\frak v+\frac{\mathcal{V}_{\textrm{SV}}'(\frak{v}+q_0)}{\mathcal{V}_{\textrm{SV}}'\Lambda \mathcal{V}_{\textrm{SV}}}\left( \frac{\sinh(\sqrt{\lambda}(T-t))}{\sinh(\sqrt{\lambda}T)} - \left(1-\frac tT\right)\right) \Lambda\mathcal{V}_{\textrm{SV}}\\
&+& \frac{\rho \frak{s} \xi - \zeta}{2\lambda} \left( \frac{\sinh(\sqrt{\lambda}t) + \sinh(\sqrt{\lambda}(T-t))}{\sinh(\sqrt{\lambda}T)} - 1\right)\Lambda\mathcal{V}_{\textrm{SV}}.\\
\end{eqnarray*}

This strategy enables to hedge progressively -- in fact linearly appears to be optimal -- the book of exotic options in the market model (this is the term $\left(1- \frac tT\right)q_0-\frac t T\frak v$ that goes linearly from $q_0$ to $-\frak v$ as $t$ goes from $0$ to $T$) while hedging the evolving portfolio in the stochastic volatility model (this is the term $ \frac{\mathcal{V}_{\textrm{SV}}'(\frak{v}+q_0)}{\mathcal{V}_{\textrm{SV}}'\Lambda \mathcal{V}_{\textrm{SV}}}\left( \frac{\sinh(\sqrt{\lambda}(T-t))}{\sinh(\sqrt{\lambda}T)} - \left(1-\frac tT\right)\right) \Lambda\mathcal{V}_{\textrm{SV}}$). As for our first model, we notice that we use a unique basket of vanilla options for hedging in the stochastic volatility model. The strategy also takes account of the view of the trader on the market as above. Unlike in the first problem however, the strategy associated with the trader's view has to be a round trip: this is the term $\frac{\rho \frak{s} \xi - \zeta}{2\lambda} \left( \frac{\sinh(\sqrt{\lambda}t) + \sinh(\sqrt{\lambda}(T-t))}{\sinh(\sqrt{\lambda}T)} - 1\right)\Lambda\mathcal{V}_{\textrm{SV}}$.\\

\section*{Conclusion}

In this paper, we built a framework in which exotic options are priced using a market model and where the prices of vanilla options are driven by a stochastic volatility model. Using this framework, we derived, after some simplifying approximations, the optimal  hedging strategy associated with a book of exotic options in presence of execution costs \`a la Almgren-Chriss to trade vanilla options. In the case of quadratic execution costs, we even showed that the hedging strategy could be computed in closed form.

\bibliographystyle{plain}

\end{document}